\documentclass{aa}
\newcommand {\hei}{He~{\sc{i}}}
\newcommand {\heii}{He~{\sc{ii}}}
 
\newcommand {\hi}{H\,{\sc i}} 
\newcommand {\kms}{\relax \ifmmode {\,\rm km\,s}^{-1}\else \,km\,s$^{-1}$\fi}
\newcommand {\ha}{H$\alpha$}
\newcommand {\hb}{H$\beta$}
\newcommand {\hg}{H$\gamma$}
\newcommand {\oiii}{O\,{\sc iii}}
\newcommand {\niii}{N\,{\sc iii}}
\newcommand {\siii}{Si\,{\sc iii}}
\newcommand {\ciii}{C\,{\sc iii}}
\newcommand {\hd}{HD\,108}
\newcommand {\x}{{\it XMM-Newton}}
\newcommand {\ro}{{\it ROSAT}}
\usepackage{graphicx}
\usepackage{txfonts}
\usepackage{rotating}
\usepackage{array}

\begin{document}

   \title{\hd: the mystery deepens with \x\ observations\thanks{Based on observations collected at the Observatoire de Haute-Provence (France) and with \x, an ESA Science Mission with instruments and contributions directly funded by ESA Member States and the USA (NASA).}}

   \subtitle{}

   \author{Y. Naz\'e\thanks{Research Fellow FNRS (Belgium)}, G. Rauw\thanks{Research Associate FNRS (Belgium)}, J.-M. Vreux, \& M. De Becker}

   \offprints{Y. Naz\'e \email{naze@astro.ulg.ac.be}}

   \institute{Institut d'Astrophysique et de G\'eophysique;
	Universit\'e de Li\`ege;
	All\'ee du 6 Ao\^ut 17, Bat. B5c;
	B 4000 - Li\`ege;
	Belgium 
             }
\authorrunning{Naz\'e et al.}
   \date{}

   \abstract{ 

In 2001, using a large spectroscopic dataset from an extensive monitoring 
campaign, we discovered that the peculiar Of star \hd\ displayed extreme 
line variations. This strange behaviour could be attributed to a variety of 
models, and an investigation of the high energy properties of \hd\ 
was needed to test the predictions from these models.
Our dedicated \x\ observation of \hd\ shows that its spectrum is well 
represented by a two temperature thermal plasma model with $kT_1\sim0.2$~keV 
and $kT_2\sim1.4$~keV. In addition, we find that the star does not display 
any significant short-term changes during the \x\ exposure. Compared 
to previous $Einstein$ and \ro\ detections, it also appears that \hd\ 
does not present long-term flux variations either.
While the line variations continue to modify \hd's spectrum 
in the optical domain, the X-ray emission of the star appears thus 
surprisingly stable: no simple model is for the moment able to 
explain such an unexpected behaviour.  

Thanks to its high sensitivity, the \x\ observatory has also enabled the 
serendipitous discovery of 57 new X-ray sources in the field of \hd. 
Their properties are also discussed in this paper. 

   }

   \maketitle
\section{Introduction}

Some O-type stars are still challenging astronomers many decades after 
their discovery. \hd\ is a good example of such an object. In the past, 
this star has been the target of several investigations, but apparently,
none could arrive at a consensus on the exact nature of this peculiar star.
Consequently, various theories have successively been proposed: \hd\ 
has been classified as a short-term binary (having even survived a 
supernova event according to some authors); it has been proposed to be 
a single star experiencing wind variability and/or harbouring a disc 
and jets; and finally it was suggested to be a long-term binary
(see Naz\'e et al. 2001, hereafter Paper I, and references therein).
In Paper I, we presented a 30 year campaign of optical spectroscopy 
dedicated to \hd. We showed that the behaviour of \hd\ was not that
of a classical short- or long-term SB1 binary. The star also did not
present any short-term variations. But our extensive campaign clearly 
indicated the peculiar characteristics of this star: tremendous line
variations on the timescale of decades. 
The hydrogen and \hei\ lines change from strong P Cygni 
profiles to simple absorptions, and a few other emission lines 
apparently follow the same behaviour. Comparing with all data 
available, we noted that these variations were recurrent, on a timescale
of a few decades. Such a continuous decline of \hi\ and 
\hei\ lines was recently discovered by Walborn et al. (\cite{wal03}) 
in another Of?p star, HD\,191612. The line variations of this star
are strikingly similar to \hd, but occur on a shorter timescale. 
In this context, we note that HD\,191612 has now returned to a 
high emission state: a spectrum of this star covering the \ha\ line, taken
in October 2003 at the Observatoire de Haute-Provence, is almost identical
to that of August 1997 presented by Walborn et al. (\cite{wal03}).

The nature of these Of?p stars is still unknown. The most 
popular models to explain their peculiar behaviour involve the presence 
of a compact companion on an elliptical orbit, or require the star to be 
a single rapid rotator (see e.g. the review in Walborn et al. \cite{wal03}). 
It was also suggested that these stars could be transition objects, 
explaining their small number (only 3 are known in our Galaxy). 
In this context, X-ray observations represent an important 
opportunity to better understand these objects, since stars
reveal at high energies the most exotic processes taking place in 
their vicinity. X-ray data are especially well suited to test the possibility 
of accretion processes linked to the presence of a compact object
or a colliding wind interaction in a binary.
We thus decided to observe \hd\ with the \x\ satellite, in the hope that
its high sensitivity could provide us with definitive answers on the nature
of \hd.

In this paper, we describe in Sect.~2 the observations 
used in this study. The optical and X-ray data of \hd\ will then
be successively analysed in Sect.~3. The remarkable sensitivity of \x\
also enabled the serendipitous discovery of many fainter sources during 
our observation. These sources detected in the field of \hd, their possible 
counterparts, their hardness ratios (HRs), their variability, 
and their spectral characteristics will be discussed in Sect.~4. 
Finally, we will conclude in Sect.~5.

\section{Observations}

\subsection{X-ray data}

\hd\ was observed with \x\ in the framework of the guaranteed time of the
Optical Monitor Consortium during revolution 494, on Aug. 21 2002, 
for approximately 35~ks. The two EPIC MOS cameras (Turner et al. \cite{tur})
were used in full frame mode, and the EPIC pn instrument (Str\"uder 
et al. \cite{str}) was operated in extended full 
frame mode. A thick filter was added to reject optical light.

\begin{figure}
\begin{center}
\end{center}
\caption{EPIC hardness ratio map of the whole field around \hd. Three 
energy bands were used to create this color image: red corresponds 
to 0.4-1.0 keV, green to 1.0-2.0 keV and blue to 2.0-10.0 keV. \hd\ 
is the bright source at the center of the field.
\label{color}}
\end{figure}

We used the Science Analysis System (SAS) software version 5.4.1. to 
reduce the EPIC data. These data were first processed through the 
pipeline chains, and then filtered.
For EPIC MOS, only events with a pattern between 0 and 12 and 
passing through the \#XMMEA\_EM filter were considered. For EPIC pn, 
we kept events with flag$=$0 and a pattern between 0 and 4. 
To search for contamination by low energy protons, we 
examined the light curve at high energies (Pulse Invariant 
channel number $>$10000, $\sim$E$>$10 
keV, and with pattern$=$0). A very short flare was detected and we
thus discarded the time intervals with a high energy count rate 
larger than 0.2 cts s$^{-1}$ (for EPIC MOS) and 0.64 cts s$^{-1}$ 
(for EPIC pn).  The resulting useful exposure times are 35.7~ks 
and 28.9~ks for EPIC MOS and pn, respectively. Further analysis 
was performed using the SAS v5.4.1. and the FTOOLS tasks. 
The spectra were analysed and fitted within XSPEC v11.0.1. 

\begin{figure*}
\begin{center}
\includegraphics[width=15cm]{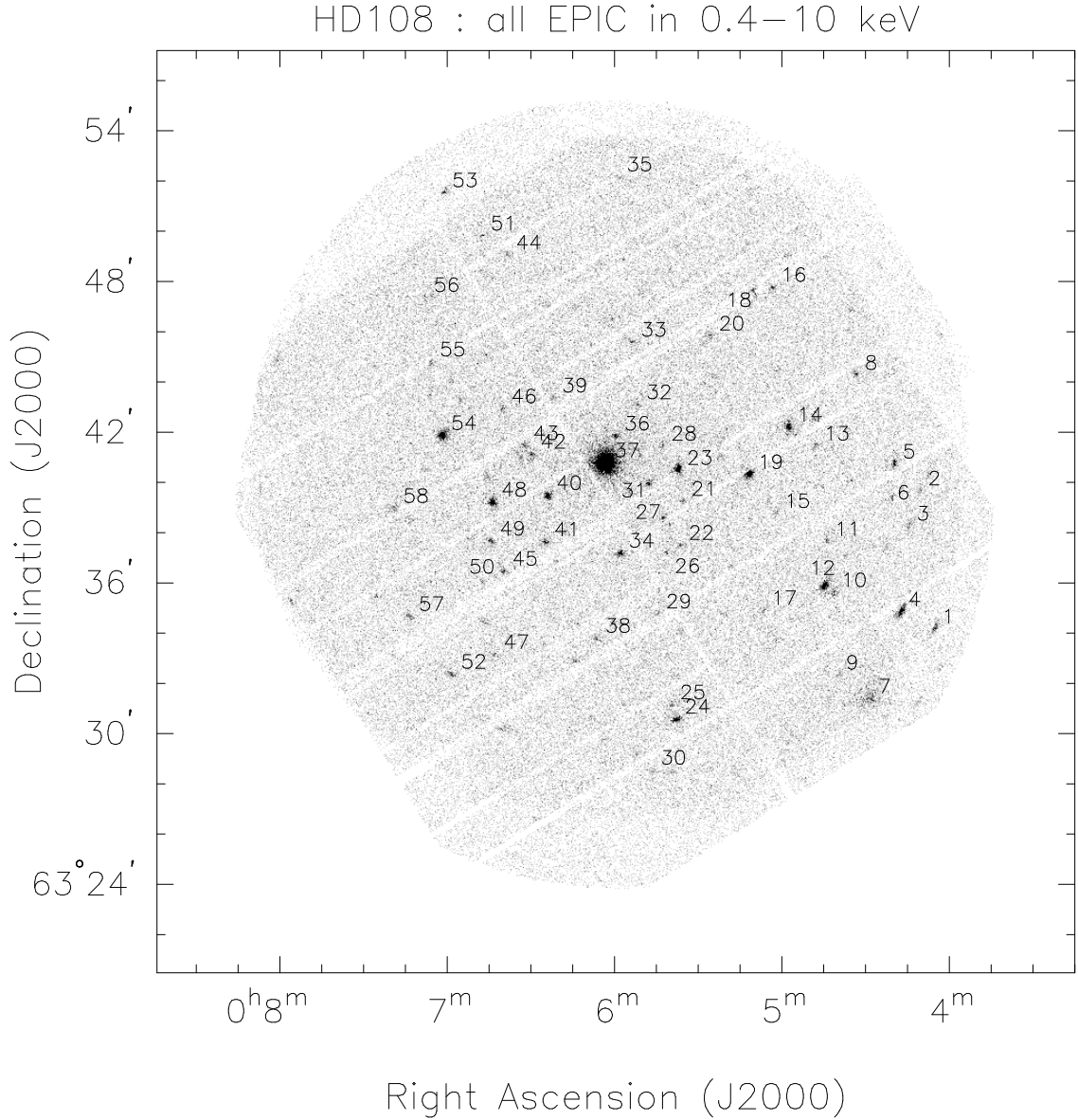}
\end{center}
\caption{EPIC image of \hd\ in the range 0.4-10 keV. The detected sources 
are labelled. The image has been binned by a factor of 50, to obtain a 
pixel size of 2.5\arcsec. 
\label{totfield}}
\end{figure*}

The first two figures show images of the field from the combined EPIC 
instruments. Figure \ref{color} presents a three colour image, 
in which numerous hard X-ray point sources are clearly 
seen. Figure \ref{totfield} gives the identification of the point 
sources detected in the field (see Sect. 4 and Table \ref{crate}).  

The RGS data were processed through the `rgsproc' pipeline SAS
meta-task. This routine determines the location of the source spectrum 
on the detector (see e.g.\ den Herder et al.\,\cite{RGS}), and performs 
the extraction of the source spectrum as well as of the background 
spectrum from a spatially offset background region. Finally, the task 
computes tailored response matrices for the first and second order RGS 
spectra.

\subsection{Additional optical data}

In 2001, 2002 and 2003, we continued our extensive spectroscopic survey 
of \hd\ with the Aur\'elie spectrograph attached to the 1.52\,m telescope 
of the Observatoire de Haute-Provence (OHP). The detector was a 
2048$\times$1024 
CCD EEV 42-20\#3, whose pixel size is (13.5$\mu$m)$^2$. The exact 
wavelength ranges and mean S/N are given in Table \ref{tab obs}, together 
with the dispersion. All the data were reduced in the standard way using 
the MIDAS software developed at ESO. The spectra were normalized by 
fitting splines through carefully chosen continuum windows, the same 
as the ones used in Paper I. 

\begin{table}[htb]
\caption { \label{tab obs} Summary of our CCD spectroscopic observations. $N$ indicates the total number of spectra obtained during a run.} 
\begin{center}
\begin{tabular} {l r r r r c c}
\hline
 Date & Wav. range & $N$ & S/N & Disp.& $R$\\
& & & &(\AA\,mm$^{-1}$)\\
\hline\hline
Sep. 2001& 3680-4130& 2&160& 16&7000\\
Sep. 2001& 4080-4540& 2&160& 16&7000\\
Sep. 2001& 4450-4900& 6&200& 16&7000\\
Sep. 2001& 6350-6760& 2&360& 16&11000\\
Sep. 2002& 4450-4905& 2&250& 16&7000\\
Sep. 2002& 6350-6760& 1&260& 16&11000\\
Oct. 2003& 4450-4915& 5 & 40-350& 16& 7000\\
Oct. 2003& 6350-6750& 1& 300& 16 & 11000\\
\hline
\end{tabular}
\end{center}
\end {table}

\section{\hd}

\subsection{Latest optical data}

\begin{figure}
\includegraphics[width=8cm]{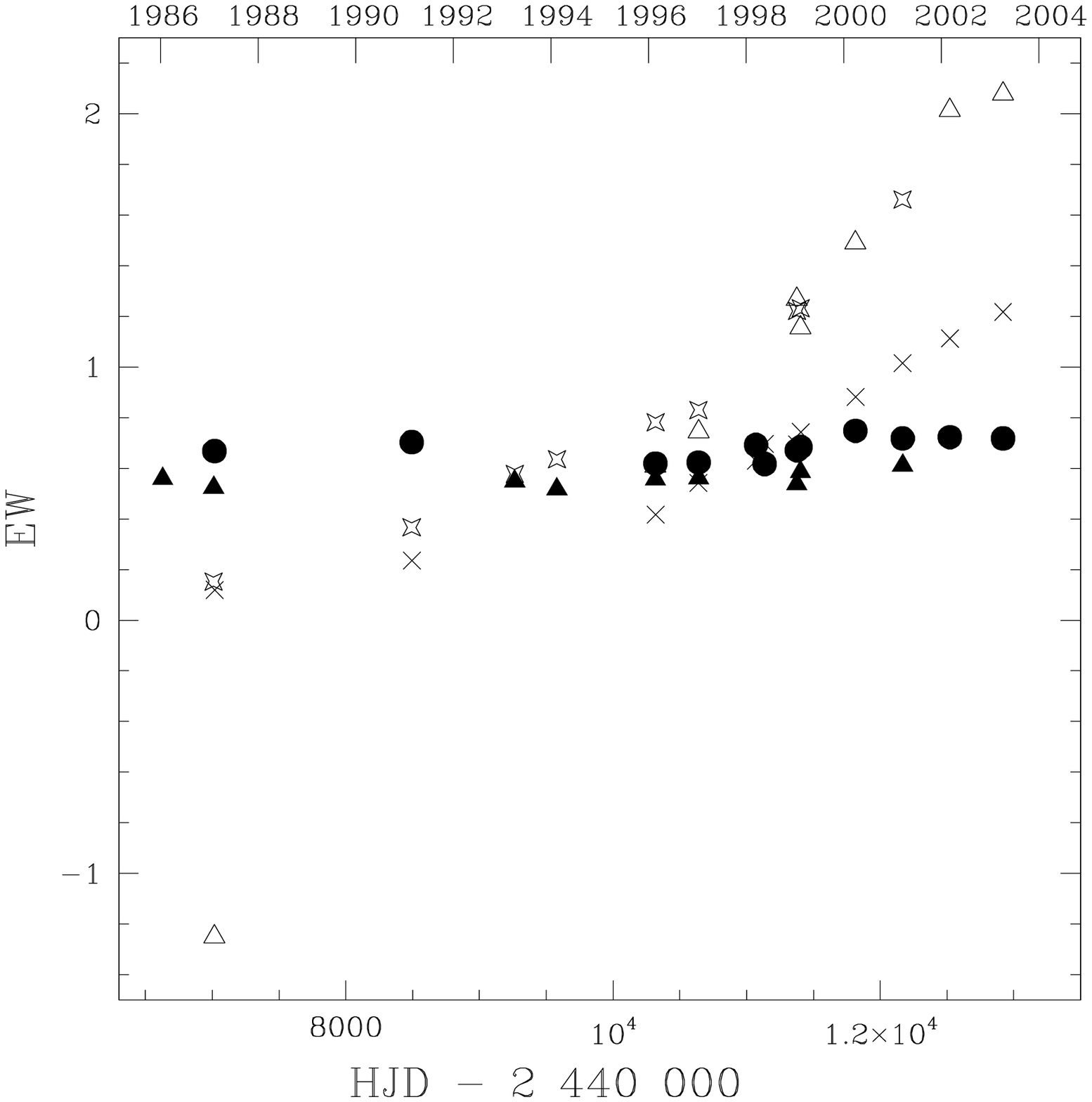}
\caption{ \label{ew} Average EWs for all observing runs: 
\heii\ $\lambda$\,4200 is represented by filled triangles, H$\gamma$ by 
distorted squares, \hei\ $\lambda$\,4471 by crosses, \heii\ $\lambda$\,4542 
by filled circles, and H$\beta$ by open triangles. The errors 
(evaluated from the data dispersion) are very 
small, roughly the size of the symbols used in the figure.}
\end{figure}

After a reduction consistent with that of the data from previous years 
(see Paper I), we found that \hd\ had still not reached its minimum 
emission stage. The \hi, \hei, \siii, 
and \ciii\ lines continued to decline. 
The average equivalent 
widths\footnote{In Table 5 of Paper I, the EW of \hb\ should 
be read $-1.25$\AA\ in 1987 and 1.49$\pm$0.07\AA\ in 2000. 
} (EWs) were 1.66$\pm$0.02\AA\ in 2001 for the \hg\ line; 
1.02$\pm$0.03\AA\ in 2001, 1.11$\pm$0.01\AA\ in 2002 and 
1.22$\pm$0.04\AA\ in 2003 for \hei\ $\lambda$4471; 
0.72$\pm$0.01\AA\ in 2001,
 0.72$\pm$0.02\AA\ in 2002 and 0.72$\pm$0.04\AA\ in 2003 for \heii\ 
$\lambda$4542; 2.01$\pm$0.03\AA\ in 2002 and 2.08$\pm$0.12\AA\ in 2003
for \hb.
We show in Fig. \ref{ew} the evolution of the EWs
with time: the changes are striking. We also note that the \hb\
line may now also present variations on a shorter timescale: in 2001
and 2003, the absorption apparently increased in the last spectra of the runs. The spectral type of the star has also continued to change through the
years, as we had predicted in Paper I: it was O8Ifpe in 2001-2002 and 
O8.5fpe in 2003.

\begin{figure*}
\begin{center}
\includegraphics[width=17cm]{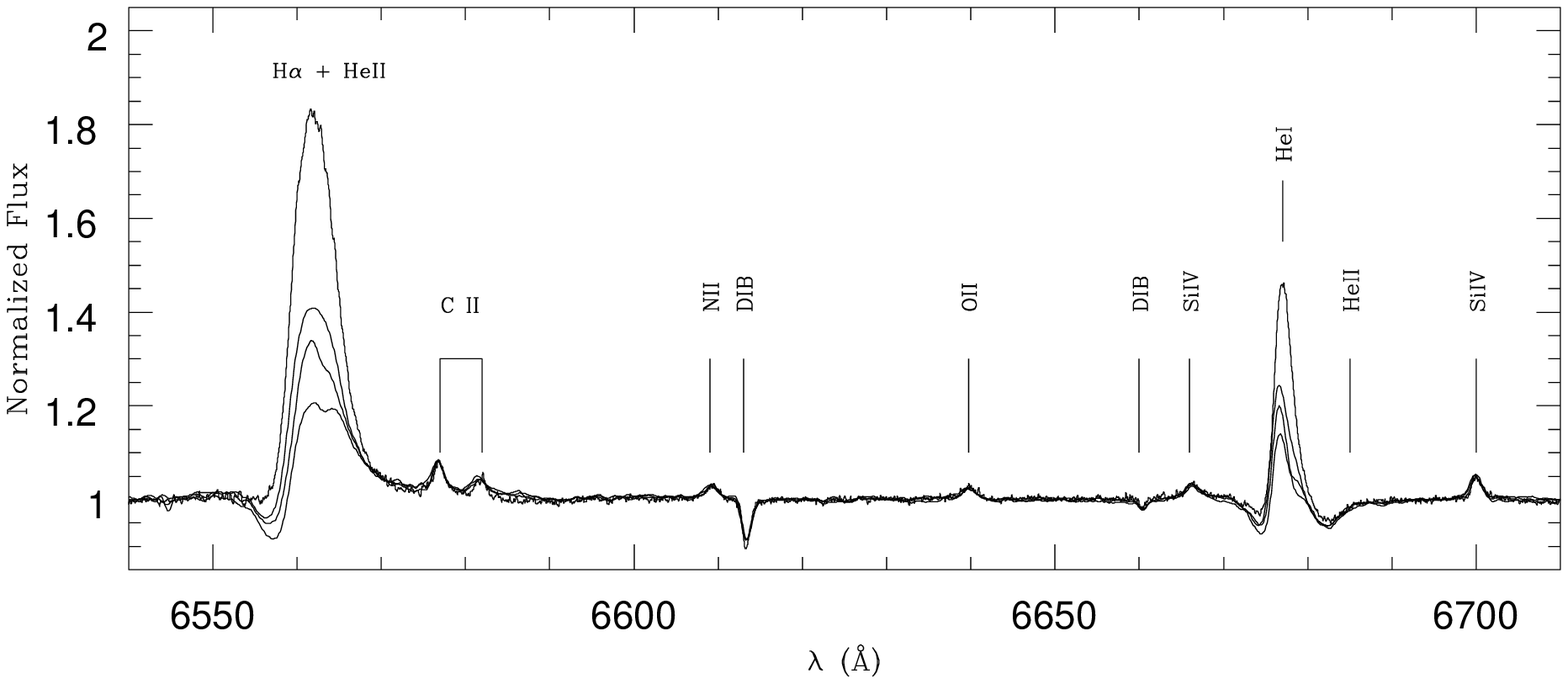}
\end{center}
\caption{ \label{ha} Evolution of the red spectrum of \hd\ with time. 
In the 1997 data, the \ha\ line was in pure emission while in the 21st
century data, it exhibits a clear P Cygni profile, whose emission 
component decreases with time. }
\end{figure*}

In 1997, 2001, 2002 and 2003, we also 
obtained spectra of \hd\ in a region 
containing the \ha\ line. This line too underwent dramatic variations, 
as is shown in Fig. \ref{ha}: its EW changes from 
-5.34\AA\ in 1997 to -1.37\AA\ in 2003! The profile is now asymetric, 
with a small absorption in the blue wing. We can use this \ha\ EW
to estimate the mass-loss rate of the star (Lamers \& Leitherer 
\cite{lam93}). But first, we have to correct the observed EW 
for photospheric \hi\ and \heii\ absorptions. To this aim, we 
decided to use the observed \ha\ EW of O8I and O8.5I stars listed in
Conti (\cite{con}). However, the only value available in Conti's 
paper, 2.91\AA, is that of HD\,167771, which is currently known 
to be a binary star. Since this could affect the determination of the EW 
of its \ha\ line,
we rather use the average EW as measured by Conti (\cite{con}) 
on the spectra of single O7-9.5I stars (HD\,34656, HD\,163800, 
HD\,16429, HD\,218915). This value is 
3.06\AA. Using the method outlined in Lamers \& Leitherer (\cite{lam93}) 
with the physical parameters appropriate to \hd\ (see Sect. 3.2.1), 
an observed V magnitude of 7.4 and color (B-V) of 0.18~mag 
(see Paper I), we then find that the mass-loss rate $\log(\dot M)$ 
varies from -5.4 to -5.5 between 1997 and 2003, whatever the 
photospheric EW correction chosen. These values of $\dot M$ 
are in between the extreme estimations from the literature (see 
below). Note however that the method of Lamers \& 
Leitherer (\cite{lam93}) assumes that the stellar wind is spherically
symmetric and homogeneous. The former hypothesis might not be valid 
in the case of \hd\ (see discussion below).

In 2001, we observed \hd\ in the violet range between 3680 and 
4130 \AA. No \oiii\ $\lambda\lambda$3755,3760 emission was detected. 
These lines would be seen in emission if the \niii\ $\lambda\lambda$4634,4640
emissions were produced through a Bowen mechanism involving the \oiii\ 
and \niii\ transitions at 374\AA. But since these \oiii\ lines are on
the contrary in absorption, this mechanism is clearly not at work in the
wind of \hd. 
The \niii\ emission lines are then most probably 
explained by a combination of the processes described in Mihalas 
(\cite{mih}), i.e. excitation of certain autoionising 
states of \niii\ followed by dielectronic recombinations (Mihalas 
\cite{mih}) and pumping of adequate \niii\ transitions by the 
intense continuum radiation of the star in an expanding atmosphere 
(Swings \cite{swi}).

\subsection{X-ray data analysis}

\begin{figure}[htb]
\begin{center}
\includegraphics[width=9cm]{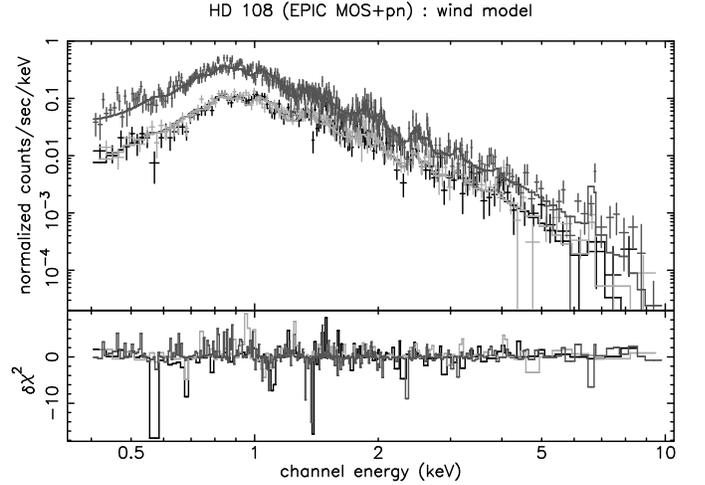}
\end{center}
\caption{The EPIC MOS and pn spectra of \hd, with the best fit 
model with wind absorption (see Table \ref{spec108}) superimposed. 
Black lines are for EPIC MOS1 data, fit, and residuals; 
light grey is for EPIC MOS2 and dark grey for EPIC pn. \label{epic}}
\end{figure}

EPIC spectra and lightcurves of \hd\ were extracted over a circular 
region centered on the star and of radius 50\arcsec\ for MOS and 
37\farcs5 for pn (because of the presence of a gap nearby).
Since a fainter source is present near \hd, we did not choose an annular
background region, but rather a nearby circle devoid of sources. 
The EPIC lightcurves did not present any significant variation. 
The spectra of each instrument were fitted separately in XSPEC, 
but as they gave the same properties, within the errors, we decided 
to fit simultaneously all EPIC data available.
We fixed the interstellar absorbing column to $N_H({\rm ISM})=3.4\times10^{21}$ 
cm$^{-2}$ (Diplas \& Savage \cite{dip}). 
The best fit was then obtained 
with a sum of two absorbed optically thin plasma models ($mekal$, Kaastra 
\cite{kaa92}) of temperatures $kT=$ 0.2 and 1.4 keV (see Fig. \ref{epic}). 
A fit by a differential emission measure model ($c6pmekl$, Sing et al. \cite{sin}) 
does not improve the quality of the fit but confirms the existence of two 
dominant temperatures. We note the presence, at $\sim$6.6~keV, of 
the Fe K-line.

\begin{table*}
\begin{center}
\caption{Spectral properties of \hd. 
The spectra were fitted simultaneously on the data from the three EPIC 
cameras and the two RGS gratings. The fluxes are in the 0.4-10.0~keV energy 
range and the emission measures are given for a distance of 2.51 kpc 
(Gies \cite{gie}). The models were of the type 
$wabs*(abs_1*mekal_1+abs_2*mekal_2)$, with the first absorbing column 
fixed to the interstellar value, $N_H({\rm ISM})=3.4\times10^{21}$ cm$^{-2}$ 
and the unabsorbed fluxes listed below were dereddened only for this column. 
The absorption components $abs_1$ and $abs_2$ are either from neutral 
gas ($wabs$) or an ionised wind (see text). The quoted errors correspond
to the 90\% confidence intervals.
\label{spec108}}\medskip
\begin{tabular}{l c c c c c c c c c } 
\hline
Abs. & Column$_1$  & $kT_1$& $EM_1$ &Column$_2$ & $kT_2$& $EM_2$ & Flux & Unabs. Flux &$\chi^2$(d.o.f.) \\ 
& 10$^{22}$ cm$^{-2}$& keV & cm$^{-3}$ & 10$^{22}$ cm$^{-2}$& keV & cm$^{-3}$ & \multicolumn{2}{c}{10$^{-13}$ erg cm$^{-2}$ s$^{-1}$}& \\
\hline
\hline
\vspace*{-0.3cm}&&&&&&&&&\\
$wabs$ & 0.51$_{0.48}^{0.53}$ & 0.20$_{0.20}^{0.22}$ & 9.4$\times$10$^{56}$ & 1.30$_{1.13}^{1.48}$ & 1.37$_{1.30}^{1.44}$ & 6.9$\times$10$^{55}$ & 6.85 & 15.7 & 1.18 (852)\\
\vspace*{-0.3cm}&&&&&&&&&\\
$wind$ & 0.54$_{0.48}^{0.57}$ & 0.26$_{0.26}^{0.27}$ & 3.7$\times$10$^{56}$ & 1.26$_{0.99}^{1.49}$ & 1.54$_{1.46}^{1.65}$ & 5.8$\times$10$^{55}$ &7.00 & 17.6 & 1.15 (852)\\
\vspace*{-0.3cm}\\
\hline
\end{tabular}
\end{center}
\end{table*}

\begin{figure}[htb]
\begin{center}
\includegraphics[width=9cm]{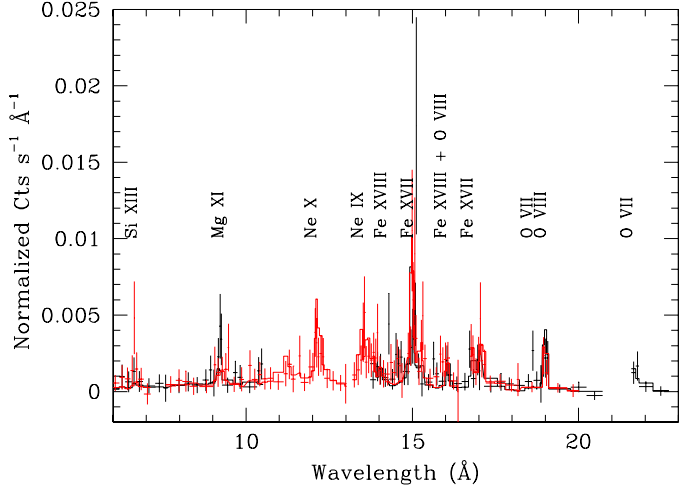}
\end{center}
\caption{The RGS1 and RGS2 spectra of \hd\ between 6 and 23\,\AA, along
with the best fit wind model of Table \ref{spec108}. The strongest lines 
are identified.\label{rgsspec}}
\end{figure}

Figure\,\ref{rgsspec} shows the first order RGS1 and RGS2 spectra of 
\hd. Unfortunately, the signal to noise ratio of individual lines 
is not sufficient to perform a detailed quantitative analysis of the 
RGS line spectrum of \hd. However, the emission lines that are 
seen in these spectra clearly indicate that the low energy ($\le$2~keV)
X-ray emission arises in a thermal plasma. We identified the strongest 
emission lines by comparison with the line list of the SPEX plasma 
code (Kaastra et al.\,\cite{kaa02}). The temperatures of maximum 
emissivity of the lines seen in the RGS spectra span a range from 
$\sim 2 \times 10^6$ to $\sim 10^{7}$\,K. These values are compatible 
with the temperatures obtained with the spectral fits of the EPIC data. 
We therefore included the RGS data in a simultaneous fit of 
all the X-ray data of \hd\ and the results are presented in Table 
\ref{spec108}. Unlike the cases of $\zeta$\,Pup 
(Kahn et al.\,\cite{Kahn}) and 9\,Sgr (Rauw et al.\,\cite{9sgr}), 
almost no lines are seen above $\sim 20$\,\AA, where the spectrum 
is heavily absorbed by circumstellar and interstellar material.

\subsubsection{Wind modelling}

The material in the stellar winds of early-type stars is 
ionised by the stellar radiation. Several authors 
(e.g. Krolik \& Kallman \cite{kro}, Waldron \cite{wal84}) have 
demonstrated that a comprehensive study of the X-ray throughput 
from massive stars requires a detailed modelling of the opacity 
from such an ionised wind. We have thus attempted to model the wind 
opacity of \hd. 

We have considered the 10 most abundant elements (H, He, C, N, O, Ne, 
Mg, Si, S, Fe) and fixed their abundances to the solar ones (Anders 
\& Grevesse \cite{and}). In our model, the collisional excitation, 
the photoionisation and the radiative and dielectronic recombinations 
determine the ionisation of the elements. Collisional excitation rates 
were taken from Voronov (\cite{vor}), and the photoionisation cross 
sections from Verner et al. (\cite{ver96}) for the outer shells and 
Verner \& Yakovlev (\cite{ver95}) for the inner shells. Since innershell 
ionisation can lead to the ejection of several electrons, we used the 
Auger yields from Kaastra \& Mewe (\cite{kaa93}) to model this 
effect for the elements considered. Radiative and dielectronic 
recombination rates were taken from Verner \& Ferland (\cite{verfer}) 
and Shull \& Van Steenberg (\cite{shu}), taking into account the 
corrections from Arnaud \& Rothenflug (\cite{arn}) for the latter reference. 

We determined the ionisation structure of the stellar wind up to 
$\sim$300 stellar radii (with a logarithmic spatial bin from 5 $R_*$ 
to 321 $R_*$), using the approximations of Waldron (\cite{wal84}) for 
the radiation field (his eq. 12) and the opacities (his eqs. 13 \& 14). 
However, we chose to fix the velocity law and the wind temperature to 
$v(r) = v_{\infty}\,(1-0.99\frac{R_*}{r})^{0.8}$ and
$T_{wind}(r)=T_{\infty}+(T_0-T_{\infty})\,(\frac{R_*}{r})^{1.9}$ (with $T_0=T_{eff}$, $T_{\infty}=$0.4$*T_0$, see Lamers \& Morris \cite{lam94}), respectively. 
The stellar parameters needed in these equations were taken from several 
sources: Howarth \& Prinja (\cite{how89}) and Peppel (\cite{pep}) give 
$R_*=17\,R_{\odot}$, $M\sim65\,M_{\odot}$ and $T_{eff}\sim40$~kK, while 
Howarth et al. (\cite{how97}) quote $v_{\infty}=1960$~\kms. In addition,
note that the stellar flux was taken from Kurucz's library of spectra 
for the above effective temperature and $\log(g)$=4.5 (the closest value 
available) and that we modelled the local emission as a simple blackbody 
at $T_{wind}$. 

There is no consensus in the litterature about the mass-loss rate 
of \hd: values from 2$\times$10$^{-7}M_{\odot}$yr$^{-1}$ (Hutchings 
\& Von Rudloff \cite{hut}) to 5$\times$10$^{-5}M_{\odot}$yr$^{-1}$ 
(Ferrari-Toniolo et al. \cite{fer}) can be found. On the other hand, 
using the \ha\ line EW as measured on our spectra from 1997-2002, 
we found an intermediate value (see above). To test the effects of 
changing mass-loss rates on the wind ionisation, we decided to run 
our model for the two extreme mass-loss rates, and in the rest of the 
paper, we will refer to these models as `wind-low' and `wind-high', 
respectively. 

\begin{figure} [htb]
\begin{center}
\leavevmode
\includegraphics[width=8cm]{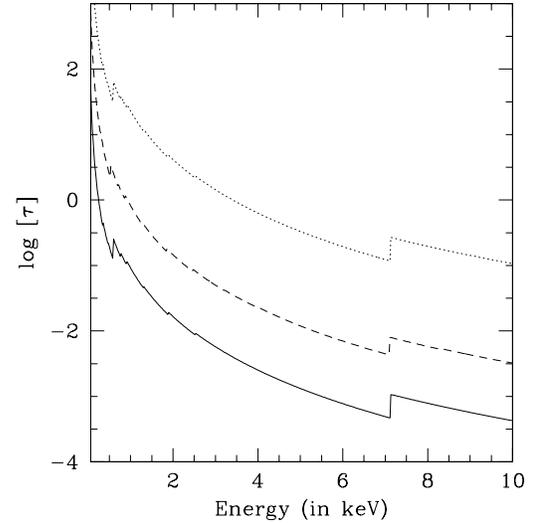}
\end{center}
\caption{Logarithm of the optical depth in the X-ray domain from 5 $R_*$ 
outwards. The solid line corresponds to $\dot M=$2$\times$10$^{-7}$M$_{\odot}$ yr$^{-1}$, and the dotted one to $\dot M=5\times$10$^{-5}$M$_{\odot}$ yr$^{-1}$. For comparison, we also show with a dashed line the optical depth of 
the neutral hydrogen column due to the interstellar medium (ISM): $N_H({\rm ISM})=3.4\times10^{21}$ cm$^{-2}$ (Diplas 
\& Savage \cite{dip}). \label{opa} }
\end{figure}

We show in Fig. \ref{opa} the resulting wind optical depth for 
both models at five stellar radii, compared to the interstellar one. 
If we scale the optical depth $\tau$ at any radius r in the wind by 
$N_{wind}(r)=\frac{X_H}{m_H}\int^{\infty}_{r} \rho(r') dr'$, we find
that the resulting opacity is very similar whatever the radius or the 
mass-loss rate considered in the model. We thus implemented a unique 
wind opacity model in XSPEC (we chose the wind-high opacity at 5 $R_*$) 
and then tried to fit the \x\ spectra using this absorbing 
model. The results are presented in Table \ref{spec108}. We see that 
within the errors, the absorbing columns are very similar compared 
to the case of neutral gas. The temperatures are slightly larger
and the Emission Measures ($EM$s) lower, compared to the case of 
the neutral gas, but the $\chi^2$ value is just slightly lower.
For an ionised wind, the two fitted columns correspond respectively 
to positions in the wind at 86 and 37 $R_*$ for the largest 
$\dot M$ (5$\times$10$^{-5}M_{\odot}$yr$^{-1}$), or 7 and 3 $R_*$, 
i.e. relatively close to the star, for the intermediate one 
($\dot M=$3.5$\times$10$^{-6}M_{\odot}$yr$^{-1}$, a value appropriate 
for the epoch of the \x\ observation if the wind is spherically 
symmetric, as assumed in our model). Note that such high columns 
could not be reached with a low value of the mass-loss rate like 
$\dot M=$2$\times$10$^{-7}M_{\odot}$yr$^{-1}$. Such a low mass-loss rate
actually produces a much lower optical depth than the interstellar matter
(see Fig. \ref{opa}). The absorption by such a wind would thus not 
lead to a significant signature at soft X-ray energies, contrary to what is 
observed: very low mass-loss rates are not compatible with the X-ray data.

\subsection{Comparison with previous X-ray observations}

The \x\ observatory is actually not the first to detect \hd\ in X-rays. 
The star had already been observed twice in the last 25 years. 
In January 1979, during a 1800~s exposure, the $Einstein$ satellite 
detected an X-ray source, 2E0003.4+6324, at the position of \hd. 
The IPC count rate for this source was 2.8$\pm$0.5 10$^{-2}$ 
cts s$^{-1}$ (2E catalog, Harris et al. \cite{har}) or 3.2$\pm$0.6  
10$^{-2}$ cts s$^{-1}$ (Chlebowski et al. \cite{chl}). Twelve years 
later, \ro\ reobserved this region during 558~s in the course of the 
All-Sky Survey. This time, the X-ray counterpart of \hd\ 
was named 1RXS J000605.0+634039 (Voges et al. \cite{vog99}) 
and presented a count rate of 6.1$\pm$1.3 10$^{-2}$ cts s$^{-1}$. 

Using the spectral properties of Table \ref{spec108}, we can predict 
the count rates expected with the IPC and PSPC\_C instruments if these properties have not changed since then. Whatever the absorption model, 
the \x\ data convert into an IPC count rate of 2.6$\pm$0.5 10$^{-2}$ 
cts s$^{-1}$ and a PSPC\_C count rate of 4.8$\pm$2.9 10$^{-2}$ cts s$^{-1}$. 
The observed count rates are thus compatible, within the errors, to what 
could be expected on the basis of the \x\ data: we observe at most a marginal 
decrease of the flux. Contrary to the optical domain, there is thus 
apparently no dramatic change at the X-ray energies on a timescale of 
decades.

\subsection{The nature of \hd}

Before we consider the various models for \hd, let us first highlight
the most important results of our \x\ investigation. 
The \x\ observations of \hd\ have revealed a thermal spectrum, of 
characteristic temperatures 0.2 and 1.4 keV. Moreover, the X-ray flux 
of \hd\ appears rather stable on both short timescales (of the order 
of the observation length) and longer ones (of the order of decades). 

To better understand the nature of the system, we can first compare 
the X-ray luminosity of \hd, corrected for the 
interstellar absorption, to its bolometric luminosity. In the \ro\ band 
(0.1-2.0 keV), we find $\log(\frac{L_X^{unabs}}{L_{BOL}})\sim-6.15$, 
whereas we would expect a value of $-6.80$ from the relations of 
Bergh\"ofer et al. (\cite{ber}). \hd\ is thus slightly overluminous 
in X-rays, with a factor of 4 between the expected and observed 
luminosities. But this difference is not as large as e.g. in some 
colliding-wind 
binaries, and our observed ratio is in fact just at the upper limit 
of the dispersion of the data of Bergh\"ofer et al. 
$L_X^{unabs}/L_{BOL}$ relations were also studied more recently 
by Moffat et al. (\cite{mof}) in the 0.5$-$10~keV energy range, 
using the $Chandra$ data of NGC\,3603:
\hd\ is then again amongst the stars at the upper limit of the dispersion.

The emission measure ($EM$) of the entire wind (assuming spherical symmetry) 
is $EM=\int n_e n_H dV\sim \frac{4\pi X_H^2}{m_H^2} \int_{R_*}^{\infty}\rho^2r^2dr$.
For the intermediate  mass-loss rate $\dot M=$3.5$\times$
10$^{-6}$M$_{\odot}$ yr$^{-1}$ and with $\rho(r)=\frac{\dot M}{4\pi r^2v(r)}$, 
we find $EM=3.9\times10^{59}$ cm$^{-3}$. 
By comparing this value to the fitted $EM$s (see Table \ref{spec108}),
we can derive the volume filling factor of the hot gas. It is approximately
$10^{-3}$ for the lower temperature component and $\sim2\times$10$^{-4}$ 
for the higher temperature one. These filling factors are in agreement 
with values derived by Kudritzki et al. (\cite{kud}) or Feldmeier et al. 
(\cite{fel}) for other early-type stars. However, we may note that for 
the largest $\dot M$ considered above, the factor would be 200 times 
lower and become, then, more difficult to reconcile with current 
X-ray emission theories and observations.

\subsubsection{A binary system with a compact object}

In Paper I, we had proposed that \hd\ could be an eccentric binary 
consisting of the Of star and a compact object with an orbital period 
of several decades. In such a system, the compact object 
could accrete matter from its companion near periastron. At that moment, 
if the X-ray luminosity produced by wind accretion becomes sufficient, 
the ionisation structure of the O-star wind
could be severely altered, and e.g. elements like hydrogen and helium 
could become completely ionised. The wind emissions of \hi\ and \hei\ 
would thus disappear, thereby explaining the observed optical line 
variations. In such a model, we would expect in the \x\ observation 
(1) a large X-ray luminosity capable of affecting the ionisation 
of the wind and (2) large variations of the X-ray flux, on orbital 
timescales, correlated with the modulation of the optical emissions. 
However, the \x\ observation revealed a completely different 
behaviour.
The ISM-corrected X-ray luminosity measured by \x\ is rather modest 
($L_X^{unabs}\sim10^{33}$ erg s$^{-1}$ in the 0.4-10~keV energy range) 
and appeared stable compared to previous X-ray observations 
made in the last 20 years. 
It is true that the ionisation of the wind can alter its radiative
line driving, thereby modifying the mass-loss rate and thus probably the X-ray 
emission, but a constant X-ray luminosity could then only be explained 
by a delicate  equilibrium between the changing wind absorption and the 
varying emission measure, which is unlikely. 

Let us consider what would be required to produce a change in the 
ionisation of the stellar wind as suggested by the optical data. 
Tarter et al. (\cite{tar}) found that the ionisation equilibrium 
in a stellar wind illuminated by an X-ray source depends mainly on 
the parameter $\xi=\frac{L_X}{n r_X^2}$, where $L_X$ is the X-ray 
luminosity, $n$ the local number density, and $r_X$ the distance to 
the X-ray source. Hatchett \& Mc Cray (\cite{hat}) futher introduced 
a dimensionless parameter $q=\xi\frac{n_X D^2}{L_X}$, where $n_X$ is 
the number density at the position of the X-ray source and $D$ 
the orbital separation between the primary star and the X-ray 
source. Interestingly, $q$ is essentially set by the velocity law 
of the stellar wind. The changes in the intensity of \hi\ and \hei\ 
emissions in the optical spectrum of \hd\ suggest that the properties 
of the wind are changing over a large volume. Therefore, the 
results of Hatchett \& McCray (\cite{hat}, their Figs. 1 \& 2) 
indicate that $q$ should be about $\sim1$. Considering a number 
of different situations, 
Kallman \& McCray (\cite{kal}) found that $\log(\xi)$ at the 
transition between neutral and ionised H and He should be in the 
range 1.0 to 1.5 depending on the wind density and the spectral 
properties of the X-ray source. These considerations suggest that $\frac{\xi}{q}=\frac{L_X}{n_X D^2}\sim10-30$. 

On the other hand, the X-ray luminosity due to wind accretion 
onto a neutron star can be estimated from Davidson \& Ostriker 
(\cite{dav}) as a function of the orbital separation $D$. Although 
we have shown that the X-ray spectrum of \hd\ can be described 
by an optically thin thermal plasma model, and this is not the 
kind of spectral shape expected for an accretion-powered X-ray 
emission, let us assume for the moment that the intrinsic $L_X$ 
of \hd\ could entirely be assigned to the accretion by the compact 
companion. After dereddening the spectra by both interstellar 
and circumstellar absorptions, we get $L_X\sim0.75-1.5$ 10$^{34}$ 
erg s$^{-1}$ in the 0.4-10.~keV energy range. These values 
correspond to orbital separations\footnote{These orbital separations
correspond to orbital periods of 16$-$25~days if the orbit was circular
(although this is unlikely to be the case for \hd). } 
of $6-8 \, R_*$ for $\dot M=$3.5$\times$ 
10$^{-6}$M$_{\odot}$ yr$^{-1}$, and this results in 
$\frac{\xi}{q}=\frac{4\pi L_X 1.3 m_H v(r)}{\dot M}\sim0.16-0.32$. 

The intrinsic 
X-ray luminosity of \hd\ as determined with \x\ is thus probably 
not sufficient to significantly alter the ionisation of the stellar 
wind and hence trigger the changes seen in the optical spectrum. 
We note however that a low X-ray luminosity does not completely 
rule out the presence of a compact companion, since the wind 
accretion process can be inhibited under certain circumstances 
(Stella et al. \cite{ste}), but that, if present, the physical conditions
are such that this companion cannot explain the optical variations.


\subsubsection{A confined wind model}

A second type of model proposed is a rapid rotator with a shell and/or disc 
and/or jets (e.g. Walborn et al. \cite{wal03}). If a disc is present 
in the system, it can only be seen nearly face-on: no double-peaked 
profiles have ever been reported for \hd\ and spectropolarimetric 
studies also favor a face-on geometry. Since there is no modulation 
of the optical data on timescales of a few days that could be 
associated with the rotation of the star, the rotation and disc axes 
need to be nearly aligned. In such a system, long-term optical 
variations could result from precession effects e.g. in a wide 
binary, but no change in the width of the lines, and hence of 
$v \sin i$, is detectable in our spectra. 

The formation of a confined outflow could be linked to the presence 
of a strong magnetic field confining the wind towards the equatorial 
regions of the star. Such 
a magnetically confined wind could provide a straightforward 
explanation for the hard X-ray component seen in the EPIC spectra 
of \hd. Models of such confined winds predict a strong 
shock at the interface between the winds that are deviated from 
both hemispheres and the equatorial disc (see e.g. Babel \& Montmerle 
\cite{bab}, ud-Doula \cite{ud}). A change in the overall mass-loss 
rate could easily account for the changes of the intensity of optical 
recombination emission lines such as \ha\ but would also imply a 
change of the total wind emission measure proportional to 
$\Delta \dot M^2$ which should then lead to a significant correlated 
variation of the X-ray luminosity that is not observed.

\subsubsection{Other models and summary}

Another model associates \hd\ with transition objects (Paper I, 
Walborn et al. \cite{wal03} and references therein) such as Ofpe/WN9
stars or Luminous Blue Variable stars (LBVs). No particular 
characteristics in the \x\ data argue in favor of such a model, 
but these data do not provide arguments 
against either. However, as in the previous model, it is difficult to 
understand why a tremendous change in the wind structure 
where emission lines arise is not correlated with a similar dramatic 
variation at the wind position where the X-ray emission takes place.  

Finally, we note that the second plasma temperature ($kT_2=1.4$~keV) 
is rather high. Most single O-type stars display thermal X-ray emission 
with $kT<1$~keV and higher temperature emission is often attributed 
to a colliding wind phenomenon in a binary system. 
A colliding wind scenario in a wide (eccentric) 
O+O binary could also account for the slight X-ray overluminosity of \hd.
However, it is not clear how the changes in the optical spectrum could 
be related to this scenario. Also, we emphasize that there is no 
obvious evidence for such a binary system in the 
spectrum of \hd.\\

All models are now challenged by the \x\ observations. Their context 
does not permit us to understand how the large variations seen in the 
optical domain could lead to a stable X-ray luminosity. This apparent 
dichotomy could well be explained by a difference in the formation region 
of the X-ray emission and the \hi/\hei\ emissions, e.g. X-rays could 
arise in a disc while the emission lines could be formed in a jet, but it 
is difficult to reconcile this with the fact that a change in any physical
parameter of the star would affect only one of the two features. 
For example, an increase in the mass-loss rate could only lead to 
a stable X-ray luminosity if the resulting additional X-ray absorption 
and emission of the wind compensate exactly. 
Such a delicate balance between emission and absorption was invoked 
by Owocki \& Cohen (\cite{owo}) to explain the empirical $L_X/L_{BOL}$ 
relation of O-type stars. This formalism holds for a mass-loss driven by
radiation pressure. However, since no large changes of the magnitude 
of \hd\ have been reported, it seems unlikely that a change in the 
mass-loss rate of the star could be triggered by a change in the overall 
luminosity and it seems therefore excluded that the mechanism of
Owocki \& Cohen could explain the lack of X-ray variability of \hd.
More than ever, \hd\ is a puzzle for astronomers. 

\section{X-ray sources in the field}

\subsection{List of sources}

The X-ray data of \hd\ also reveal a lot of additional point sources 
in the field surrounding the O star, as can be appreciated in Fig. 
\ref{color}.  To study these discrete X-ray sources, we applied 
the detection metatask {\it edetect\_chain} simultaneously to the data 
from the three EPIC cameras. We used three energy bands: S=0.4-1.0 keV, 
M=1.0-2.0 keV and H=2.0-10.0 keV. We eliminated the false detections 
mainly due to background fluctuations by rejecting the sources with 
a likelihood $<5$ in any detector and/or a combined likehood 
$<30$. This procedure left 58 point sources in 
the field. They are shown in Fig. \ref{totfield} and their properties 
are presented in Table \ref{crate}, by order of increasing RA. 

Their background-subtracted and vignetting-corrected count rates in 
each of the EPIC cameras are also presented in this table. Note that
the SAS detection tasks do not apply a correction for out-of-time events or 
dead times. For the EPIC pn data used in extended full frame mode, 
this implies an overestimate of the exposure time (and thus an 
underestimation of the count rates) by 2.3\%. Since this amount is well 
within the errors, we choose not to correct the pn count-rates 
provided by {\it edetect\_chain}. The sensitivity limit is $\sim$0.6~cts 
s$^{-1}$ for EPIC MOS data and $\sim$1.6~cts s$^{-1}$ for EPIC pn data.
The hardness ratios for the pn data 
are also indicated in Table \ref{crate}. The hardness ratios are defined as
$HR1=(M-S)/(M+S)$ and $HR2=(H-M)/(H+M)$, where $S$, $M$, and $H$ 
correspond to the count rates in the 0.4-1.0, 1.0-2.0 and 2.0-10.0 
keV band, respectively. As was already visible from Fig. \ref{color},
most of the sources are very hard, probably due to the large absorbing
column in the Galactic plane where \hd\ lies ($l_{\rm II}=117.93^{\circ}$, 
$b_{\rm II}=+1.25^{\circ}$). 

\subsection{Identification}

We have compared our source list to the \ro\ detections in the same 
field. Four catalogs are available:  the WGA catalog (White et al. 
\cite{wga}, PSPC) and three catalogs from the \ro\ teams (1RXH and 
2RXP, \ro\ consortium \cite{rxh}a and \cite{rxp}b; 1RXS, Voges et 
al. \cite{vog99} \& \cite{vog00}). Only one source,
1RXS J000605.0+634039, was previously known in the field. It is at 
13\farcs7 from XMMU J000603.2+634046 (\#37) and at 13\farcs3 from \hd.
Although this distance is slightly larger than the positional error 
of the \ro\ source, 11\arcsec, there is little doubt that the \ro\
source corresponds to \hd, since there is no other bright \x\ source 
in the region.

To search for optical counterparts, we have cross-correlated our source 
list with the Simbad catalog, the 2MASS All-Sky Data Release, the USNO 
B1.0 and the 
GSC 2.2 catalogs. To determine an optimal radius of correlation, 
we adopted the method of Jeffries et al. (\cite{jef}) as used in
Rauw et al. (\cite{rau}), i.e. fitting the cumulative distribution 
of the number of detected sources as a function of the correlation radius 
by expression (1) of Jeffries et al. (\cite{jef}). The optimal correlation 
radius\footnote{Strictly speaking, a unique correlation radius cannot 
be used with \x\ data, since the point spread function is degrading at
large off-axis angles. The radius found here should be considered as 
a weighted average.}, i.e. the radius that includes the maximum of 
true correlations 
and the minimum of spurious correlations, is found to be around 2\arcsec.
We list the counterparts to the X-ray sources found this way in Table 
\ref{crate}, together with the separation between the X-ray and the 
visible sources, and some basic information about the visible sources. 

Source XMMU J000428.5+633121 (\# 7) is actually extended,
as can be seen from comparison with the shape of the neighbouring 
sources (see Figs. \ref{color} and \ref{totfield}). This source is 
not associated with any GSC/2MASS/USNO counterpart: with its rather large 
HR, it may be an extragalactic source. 

The source XMMU J000643.6+633912 
(\# 48) corresponds to BD+62$^{\circ}$2368, an F7V star. If we compare 
the observed colors of this star (with $B$, $V$ listed in Simbad and $J$, 
$H$, $K$ from the 2MASS catalog) to typical colors of an F7V star 
(Schmidt-Kaler \cite{sch}, Kenyon \& Hartmann \cite{ken}, Bessell \& Brett
\cite{bes}), we can estimate 
its reddening and its probable distance. Using the reddening law from 
Cardelli et al. (\cite{car}) with $R_{\rm V}=$3.1, we found an 
$E({\rm B}-{\rm V})$ in the range 0.02-0.07, and a distance of 181-193~pc.

The source XMMU J000623.9+633927 (\#40) also correlates with a 
bright star, BD+62$^{\circ}$2365, but no spectral type has been quoted 
for this star in the literature so far. 
However, its 2MASS colors and a visible spectrum obtained at
OHP in Oct. 2003 apparently favor a G2III spectral type for this star.
In a similar manner as above, we derived for this object a reddening 
$E({\rm B}-{\rm V})\sim0.18$ and a distance of 540~pc.

On the other hand, several sources listed in Table \ref{crate} present 
rather high HRs. This result does not appear completely surprising 
in the light of the $\log(N)-\log(S)$ relations of Mushotzky et al. 
(\cite{mus}) and Giacconi et al. (\cite{gia}), since these relations 
predict that a large fraction of the detected sources should 
correspond to extragalactic objects even accounting for a total
Galactic column density of $\sim$0.5$\times10^{22}$ cm$^{-2}$ along the 
line of sight\footnote{We estimated this value from the extinction map of
Schlegel et al. (\cite{schle}). These authors  caution that their maps 
may not be accurate for Galactic latitudes below $|b|\simeq+5^{\circ}$,
but they can nevertheless provide a first estimate of the total extinction.}.
Finally, there are a few known early-type stars in the field in addition to 
\hd\ and EM* CDS 1: BD+62$^{\circ}$2362 (B7III), ALS 6022 (B9III), ALS 6028 
(B), EM* CDS 3 (B),  BD+62$^{\circ}$2369 (A5V) and BD+63$^{\circ}$2105 (A9II).
None of them is detected in X-rays, but such late B and A stars are 
usually not bright X-ray emitters. We may also note the non-detection 
in X-rays of the variable star SV*SVS1455 and of the relatively bright
ALS 6011 (F5I).

\subsection{Variability}

We have analysed the lightcurves of the 8 brightest point sources (i.e. with at 
least 50 cts in EPIC MOS and 100 cts in EPIC pn in the 0.4-10.0 keV range). 
The count rates in each bin were background-subtracted using annuli around
the sources or close-by circles. The effective time bin lengths were calculated 
by taking into account the `good time intervals' defined in Sect. 2. The 
resulting lightcurves were analysed by Kolmogorov-Smirnov and $\chi^2$ tests. 
We also ran a modified probability of variability test (Sana et al. 
\cite{san})  
on these sources' event lists. In the \hd\ field, no source was found 
significantly variable over the duration of our observation.

\begin{sidewaystable*}
\begin{center}
\caption{Characteristics of sources in the field of \hd. The hardness ratios refers to EPIC pn, and the counterparts are within 2\arcsec\ of the X-ray sources. For each catalog, the column `Nr' shows the number of counterparts 
lying in this 2\arcsec\ region, and `d' gives the distance in arcseconds 
between the X-ray and the optical source, if unique. 
The quoted error bars on the count rate represent $\pm1\sigma$ poissonian
standard deviation.
 \label{crate}}\medskip
\tiny
\begin{tabular}{l l | c c c c c | c c c c  | c c c c c | c c } 
\hline
\#& Name& Crate MOS1& Crate MOS2& Crate PN & HR1 & HR2 & \multicolumn{4}{c|}{GSC} & \multicolumn{5}{c}{2MASS}& \multicolumn{2}{|c}{USNO}\\ 
& & 10$^{-3}$ cts s$^{-1}$ & 10$^{-3}$ cts s$^{-1}$ & 10$^{-3}$ cts s$^{-1}$ & & & Nr & d(") & R & B & Nr & d & J & H & K & Nr & d(") \\
\hline\hline
1& XMMU J000405.7+633408  &  & & 23.5$\pm$2.1 & 0.61$\pm$0.08 & -0.04$\pm$0.10& 0 & & & & 0 & & & & &0& \\
2& XMMU J000410.6+633939  &  1.02$\pm$0.45 & 1.83$\pm$0.58 & 3.21$\pm$0.94 & 0.89$\pm$0.21 & -0.17$\pm$0.32& 0 & & & & 0 & & & & &0 & \\
3& XMMU J000414.7+633815  &  2.61$\pm$0.62 & 2.35$\pm$0.62 & 3.27$\pm$0.99 & 0.68$\pm$0.35 & 0.08$\pm$0.30& 0 & & & & 0 & & & & &0 & \\
4& XMMU J000417.7+633449  &  18.6$\pm$1.7 & & 37.1$\pm$2.5 & 0.95$\pm$0.07  & 0.55$\pm$0.05 & 0 & & & & 1&1.7 & 17.9 & 16.2 & 15.2 &0 & \\
5& XMMU J000419.7+634042  &  4.25$\pm$0.71 & 3.96$\pm$0.67 & 9.29$\pm$1.17 & 0.59$\pm$0.19 & 0.33$\pm$0.11& 0 & & & & 0 & & & & &0 & \\
6& XMMU J000420.8+633920  &  1.68$\pm$0.55 & 1.84$\pm$0.51 & 3.37$\pm$0.77 & 0.83$\pm$0.37 & 0.35$\pm$0.20& 0 & & & & 0 & & & & &0 & \\
7& XMMU J000428.5+633121$^a$  &  & & 18.9$\pm$2.5 & 0.72$\pm$0.12 & -0.20$\pm$0.15 & 0 & & & & 0 & & & & &0 & \\
8& XMMU J000433.3+634415  &  0.95$\pm$0.36 & 4.09$\pm$0.65 & 4.30$\pm$0.82 & 0.93$\pm$0.18 & 0.21$\pm$0.17& 0 & & & & 0 & & & & &0 & \\
9& XMMU J000440.1+633219  &  1.85$\pm$0.65 & 1.53$\pm$0.58 & 2.24$\pm$0.92& -0.76$\pm$0.47 & 0.82$\pm$0.37& 1& 0.9 & 14.4 & 15.9 & 1&0.9 & 12.7 & 12.1 & 11.9 & 1& 0.7 \\
10& XMMU J000441.8+633536  &  1.98$\pm$0.45 & 1.35$\pm$0.37 & 6.45$\pm$0.89 & -0.35$\pm$0.13 & -0.35$\pm$0.29& 1& 0.6 &14.6 & 16.2 & 1& 0.3 & 13.3 & 12.8 & 12.6&1& 0.8 \\
11& XMMU J000444.1+633738  &  1.25$\pm$0.33 & & 3.58$\pm$0.59 & -0.04$\pm$0.16 & -1.00$\pm$0.23& 1& 0.7 &14.5 & 15.2& 1&0.6  & 13.3 & 12.8 & 12.7&1& 0.9 \\
12& XMMU J000445.1+633550  &  9.98$\pm$0.95 & 10.1$\pm$1.0  & 20.6$\pm$1.5  & 0.88$\pm$0.09 & 0.46$\pm$0.06& 0 & & & & 0 & & & & &0 & \\
13& XMMU J000448.0+634126  &  1.14$\pm$0.34 & 1.55$\pm$0.39 & 2.28$\pm$0.59 & -0.14$\pm$0.44 & 0.63$\pm$0.22& 0 & & & & 0 & & & & &0 & \\
14& XMMU J000457.6+634211  &  3.65$\pm$0.47 & 5.27$\pm$0.60 & 12.4$\pm$1.0  & 0.83$\pm$0.08 & 0.04$\pm$0.08& 0 & & & & 1&1.1 & 16.8 & 15.5 &14.8 &0 & \\
15& XMMU J000502.1+633844  &  & & 1.74$\pm$0.52 & 0.56$\pm$0.36 & 0.08$\pm$0.30& 0 & & & & 0 & & & & &0 & \\
16& XMMU J000503.4+634744  &  1.06$\pm$0.31 & 1.81$\pm$0.36 & & & & 1& 0.9 & 16.2 & 18.6& 1&0.8  & 13.5  & 12.9  & 12.7 &0 & \\
17& XMMU J000506.8+633453  &  0.86$\pm$0.36 & 1.23$\pm$0.41 & 1.90$\pm$0.59 & 0.39$\pm$0.39 & 0.19$\pm$0.30& 0 & & & & 0 & & & & &0 & \\
18& XMMU J000510.1+634733  &  2.14$\pm$0.39 & 2.42$\pm$0.41 & & & & 0 & & & & 1&1.3 & 14.9 & 14.3 & 13.9&1& 1.9 \\
19& XMMU J000511.7+634018  &  7.38$\pm$0.60 & 7.13$\pm$0.58 & & & & 0 & & & & 0 & & & & &0 & \\
20& XMMU J000525.6+634549  &  0.88$\pm$0.24 & 0.76$\pm$0.25 & 2.70$\pm$0.54 & -0.05$\pm$0.21 & -0.09$\pm$0.26& 1& 1.1 & & 12.0& 1&1.1 & 11.1 &11.0 & 11.0 &1& 1.1 \\
21& XMMU J000535.5+633914  &  0.52$\pm$0.22 & 0.90$\pm$0.23 & & & & 0 & & & &0 & & & & &0 & \\
22& XMMU J000536.5+633728  &  0.63$\pm$0.23 & 0.76$\pm$0.25 & 1.69$\pm$0.40 & 0.50$\pm$0.26 & -0.28$\pm$0.27& 0 & & & & 0 & & & & &0 & \\
23& XMMU J000537.1+634027  &  6.18$\pm$0.52 & 7.19$\pm$0.52 & & & & 0 & & & & 0 & & & & &0 & \\
24& XMMU J000537.9+633034  &  6.93$\pm$0.81 & 6.51$\pm$0.83 & 16.2$\pm$1.3 & 0.40$\pm$0.11 & 0.28$\pm$0.07& 0 & & & & 0 & & & & &0 & \\
25& XMMU J000539.6+633108  &  1.18$\pm$0.41 & 0.97$\pm$0.30 & 3.86$\pm$0.75 & -0.66$\pm$0.16 & -0.39$\pm$0.65& 1&0.4 & & 12.1& 1&0.4 & 10.8  &10.6 & 10.5 &1& 0.4 \\
26& XMMU J000541.5+633711  &  0.68$\pm$0.23 & 0.44$\pm$0.19 & 1.60$\pm$0.41 & 0.35$\pm$0.48 & 0.39$\pm$0.23& 0 & & & & 0 & & & & &0 & \\
27& XMMU J000542.6+633834  &  1.18$\pm$0.28 & 1.48$\pm$0.28 & & & & 1&0.8 &15.5&16.6 & 1& 0.5 & 14.2  & 13.9 & 13.8 &1& 0.7 \\
28& XMMU J000542.7+634125  &  0.59$\pm$0.20 & 0.48$\pm$0.19 & 1.48$\pm$0.38 & 0.51$\pm$0.53 & 0.48$\pm$0.23& 0 & & & & 0 & & & & &0 & \\
29& XMMU J000544.6+633444  &  1.22$\pm$0.30 & 0.63$\pm$0.28 & & & & 0 & & & & 0 & & & & &0 & \\
30& XMMU J000546.5+632831  &  1.25$\pm$0.57 & 1.46$\pm$0.56 & 2.99$\pm$0.94 & 0.002$\pm$0.36 & 0.18$\pm$0.36& 0 & & & & 0 & & & & &0 & \\
31& XMMU J000547.7+633955  &  1.52$\pm$0.28 & 1.24$\pm$0.26 & & & & 0 & & & & 0 & & & & &0 & \\
32& XMMU J000551.7+634304  &  0.75$\pm$0.17 & 0.56$\pm$0.20 & 1.13$\pm$0.34 & -0.67$\pm$0.25 & -1.00$\pm$1.83& 1&2.0 &12.4 & 14.3& 1&1.7  & 10.4 & 9.7 & 9.5 &0& \\
33& XMMU J000553.6+634533  &  0.90$\pm$0.20 & 0.89$\pm$0.23 & 1.62$\pm$0.44 & 0.53$\pm$0.47 & 0.33$\pm$0.25& 1& 1.0 &17.3 & 18.2& 1&0.8 & 15.7 & 15.5 & 15.5 &1& 1.3 \\
34& XMMU J000558.0+633710$^b$  &  1.48$\pm$0.28 & 3.67$\pm$0.39 & 6.21$\pm$0.61 & 0.62$\pm$0.12 & 0.01$\pm$0.10& 0 & & & & 0 & & & & &0& \\
35& XMMU J000558.7+635208  &  1.55$\pm$0.38 & 1.24$\pm$0.40 & & & & 0 & & & & 0 & & & & &0 & \\
36& XMMU J000559.5+634149  &  1.04$\pm$0.23 & 0.99$\pm$0.24 & 3.25$\pm$0.48 &0.53$\pm$0.21 & 0.00$\pm$0.15&0 & & & & 0 & & & & &0 & \\
37& XMMU J000603.2+634046$^c$  &  99.9$\pm$1.8 & 105.$\pm$2. & 333.$\pm$4. & -0.10$\pm$0.01 & -0.50$\pm$0.01& 1& 0.7 & & 7.5& 1&0.7 &  6.9 & 6.9 & 6.8  &1& 0.7 \\
38& XMMU J000606.4+633347  &  1.17$\pm$0.29 & 1.22$\pm$0.31 & 2.12$\pm$0.50 & -0.65$\pm$0.19 & -1.00$\pm$1.34& 0 & & & & 0 & & & & &0 & \\
39& XMMU J000621.9+634321  &  0.68$\pm$0.21 & 0.57$\pm$0.20 & 0.88$\pm$0.33 & 1.00$\pm$0.73 & 0.51$\pm$0.32& 0 & & & & 0 & & & & &0 & \\
40& XMMU J000623.9+633927$^d$  &  2.06$\pm$0.28 & 2.20$\pm$0.32 & 9.51$\pm$0.70 & -0.47$\pm$0.06 & -0.76$\pm$0.15& 1& 1.1 & 9.7& & 1&0.7  & 8.3 & 7.8 & 7.7&1& 0.8 \\
41& XMMU J000624.8+633737  &  1.54$\pm$0.29 & 1.33$\pm$0.27 & 3.77$\pm$0.53 & 0.72$\pm$0.31 & 0.51$\pm$0.12& 0 & & & & 0 & & & & &0 & \\
42& XMMU J000629.5+634106  &  1.02$\pm$0.26 & 1.30$\pm$0.27 & 2.93$\pm$0.53 & 0.05$\pm$0.19 & -0.25$\pm$0.23& 1&0.5 & 14.6 & 15.5& 1&0.4  & 12.8 & 12.3 & 12.2 &1& 0.2 \\
43& XMMU J000632.3+634127  &  0.68$\pm$0.23 & 0.53$\pm$0.18 & 2.47$\pm$0.48 & 0.50$\pm$0.24 & -0.17$\pm$0.21& 0 & & & & 0 & & & & &0 & \\
44& XMMU J000638.4+634902  &  0.73$\pm$0.29 & 0.67$\pm$0.27 & 2.25$\pm$0.58 & -0.16$\pm$0.23 & -0.54$\pm$0.49& 1& 1.5 & 14.2 & 16.0& 1&0.4&  13.8 & 13.2 & 13.0&3& \\
45& XMMU J000639.8+633627  &  1.55$\pm$0.31 & 1.07$\pm$0.28 & 3.33$\pm$0.52 & 0.68$\pm$0.25 & 0.36$\pm$0.13& 0  & & & & 0 & & & & &0 & \\
46& XMMU J000640.1+634255  &  0.63$\pm$0.21 & 0.74$\pm$0.21 & 2.00$\pm$0.46 & 0.32$\pm$0.28 & 0.02$\pm$0.25& 0  & & & & 0 & & & & &0 & \\
47& XMMU J000642.9+633308  &  0.67$\pm$0.33 & 1.07$\pm$0.38 & 3.39$\pm$0.69 & 0.31$\pm$0.24 & 0.14$\pm$0.21& 0  & & & & 0 & & & & &0 & \\
48& XMMU J000643.6+633912$^e$  &  2.95$\pm$0.31 & 2.55$\pm$0.31 & 13.5$\pm$0.1 & -0.76$\pm$0.04 & -1.00$\pm$0.07& 1&1.8 & & 10.9 & 1&1.7  &  9.3 & 9.1 & 9.1 &1& 1.8 \\
49& XMMU J000644.2+633739  &  1.48$\pm$0.30 & & 4.47$\pm$0.60 & 0.63$\pm$0.17 & -0.03$\pm$0.14& 0 & & & & 0 & & & & &0 & \\
50& XMMU J000647.2+633602  &  0.78$\pm$0.29 & 0.63$\pm$0.25 & 1.00$\pm$0.40 & 1.00$\pm$0.55 & -0.17$\pm$0.38& 0 & & & & 0 & & & & &0 & \\
51& XMMU J000647.5+634947  &  1.19$\pm$0.38 & 1.20$\pm$0.35 & 1.21$\pm$0.52 & 0.97$\pm$0.31 & -0.52$\pm$0.51& 0 & & & & 0 & & & & &0 & \\
52& XMMU J000658.1+633221  &  3.11$\pm$0.65 & 2.85$\pm$0.53 & 7.02$\pm$1.01 & -0.71$\pm$0.45 & 0.97$\pm$0.04& 0 & & & & 0 & & & & &0 & \\
53& XMMU J000701.3+635130  &  4.38$\pm$1.25 & 5.12$\pm$0.73 & & & & 0 & & & & 0 & & & & &0 & \\
54& XMMU J000701.7+634151  &  7.99$\pm$0.62 & 8.33$\pm$0.63 & 24.0$\pm$1.3 & 0.95$\pm$0.08 & 0.71$\pm$0.03& 0 & & & & 0 & & & & &0 & \\
55& XMMU J000705.7+634447  &  1.18$\pm$0.25 & 1.66$\pm$0.33 & & & & 0 & & & & 0 & & & & &0 & \\
56& XMMU J000707.8+634721  &  0.89$\pm$0.29 & 1.03$\pm$0.34 & 1.55$\pm$0.62 & 0.27$\pm$0.42 & -0.04$\pm$0.47& 0 & & & & 0 & & & & &0 & \\
57& XMMU J000713.2+633439  &  1.98$\pm$0.44 & 1.39$\pm$0.40 & 5.73$\pm$0.88 & 0.55$\pm$0.22 & 0.28$\pm$0.14& 0 & & & & 0 & & & & &0 & \\
58& XMMU J000718.8+633857  &  1.69$\pm$0.37 & 1.06$\pm$0.32 & 3.01$\pm$0.64 & -0.66$\pm$0.17 & -0.44$\pm$0.69& 1& 1.5 & 16.0 & 18.3& 1&1.6 & 13.2 & 12.5 & 12.3 &1& 1.6 \\
\hline
\end{tabular}
\end{center}
\tiny 
$^a$ Extended source\\
$^b$ The emission-line star EM* CDS 1 (B-type) is at 8\farcs5 from this source. \\
$^c$ = \hd\ (the X-ray position is at 0\farcs8 from the optical one given in Simbad).\\
$^d$ = BD+62$^{\circ}$2365 (spectral type G2III, see text)\\
$^e$ = BD+62$^{\circ}$2368 (spectral type F7V)\\
\end{sidewaystable*}
\normalsize

\subsection{Spectral properties}

\begin{table*}
\begin{center}
\caption{Spectral properties of the brightest sources of the \hd\ 
field-of-view. The spectra were fitted simultaneously on the data 
from the three EPIC cameras, except for \# 19 and 23, where there 
exist only MOS data. The observed fluxes are in units 10$^{-13}$ 
erg cm$^{-2}$ s$^{-1}$ and in the 0.4-10.0~keV energy range. Models 
with bad fit quality ($\chi^2>2$) are not listed. The quoted errors correspond
to the 90\% confidence intervals.
\label{spec}}\medskip
\begin{tabular}{l  c c c c | c c c c | c} 
\hline
\#& \multicolumn{4}{c|}{Absorbed $mekal$ model} & \multicolumn{4}{c|}{Absorbed power-law model}&\\
&$N_{\rm H}$& $kT$& Flux & $\chi^2$(d.o.f.) &  $N_{\rm H}$& $\Gamma$& Flux & $\chi^2$(d.o.f.) & Var. ?\\ 
& 10$^{22}$ cm$^{-2}$& keV &  & & 10$^{22}$ cm$^{-2}$& keV & & &\\
\hline
\hline
\vspace*{-0.3cm}&&&&&&&&&\\
12 & 1.27$_{0.88}^{1.76}$ & 3.29$_{2.36}^{5.16}$ & 1.06& 0.83 (74) & 1.19$_{0.82}^{1.82}$ & 2.06$_{1.76}^{2.56}$ & 1.16& 0.83 (74) &N\\
\vspace*{-0.3cm}&&&&&&&&&\\
14 & 0.64$_{0.42}^{0.90}$ & 3.43$_{2.47}^{5.01}$ & 0.60& 0.91 (65) & 0.84$_{0.58}^{1.14}$ & 2.42$_{2.01}^{2.82}$ & 0.58 & 0.90 (65) &N\\
\vspace*{-0.3cm}&&&&&&&&&\\
19 & 0.02$_{0.}^{0.14}$ & 82$_{16}^{100}$ & 1.08& 0.62 (47)& 0.$_{0.}^{0.13}$ & 1.11$_{0.88}^{1.42}$ & 1.18 & 0.60 (47) &?\\
\vspace*{-0.3cm}&&&&&&&&&\\
23 & 0.23$_{0.09}^{0.42}$ & 42$_{10}^{100}$ & 1.03& 1.04 (52) & 0.29$_{0.12}^{0.62}$ & 1.40$_{1.13}^{1.81}$ & 1.01 & 1.03 (52) &N\\
\vspace*{-0.3cm}&&&&&&&&&\\
40 & 0.61$_{0.34}^{0.79}$ & 0.23$_{0.18}^{0.35}$ & 0.13 & 0.79 (51) & 1.20$_{1.13}^{1.28}$ & 10$_{7.6}^{10}$ & 0.13& 0.96 (51) &N\\
\vspace*{-0.3cm}&&&&&&&&&\\
48 & 0.$_{0.}^{0.10}$ & 0.39$_{0.32}^{0.45}$ & 0.20& 1.47 (62) & & & & &N\\
\vspace*{-0.3cm}&&&&&&&&&\\
54 & 1.63$_{1.30}^{1.95}$ & 19$_{11}^{100}$ & 1.72& 0.74 (95) & 1.92$_{1.43}^{2.64}$ & 1.60$_{1.37}^{1.93}$ & 1.69& 0.72 (95) &N\\
\vspace*{-0.3cm}\\
\hline
\end{tabular}
\end{center}
\end{table*}

We also extracted the spectra of these brightest sources. We 
generated response matrix files (rmf) and ancillary response files (arf)
using the SAS tasks {\it rmfgen} and {\it arfgen}. The spectra were 
then binned to reach a minimum of 10 cts per channel. Finally, we analysed 
the background-corrected spectra within XSPEC. Due to strong noise at 
very low and very high energies, we have discarded energy bins below 
0.4 keV and above 10 keV. We have fitted the spectra
with an absorbed $mekal$ model or an absorbed power-law. For each source,
we have fitted separately EPIC MOS1+2 and EPIC pn data, but as they
gave similar results - within the errors - we finally fitted all three
instruments simultaneously and we list the parameters of the best-fit 
models in Table \ref{spec}. Note that XMMU J000643.6+633912 (\# 48), 
which corresponds to the F7V star BD+62$^{\circ}$2368, is only well 
fitted by a $mekal$ model, as 
could be expected for a coronal source. For the average distance of 
187~pc estimated above, the X-ray luminosity of this source is 
8.2$\times$10$^{28}$ erg s$^{-1}$, which corresponds to 
$\log(\frac{L_X^{unabs}}{L_{BOL}})=-5.1$.
A $mekal$ model also provides a good fit for 
XMMU J000623.9+633927 (\# 40), and for the distance of 
540~pc estimated above, its X-ray luminosity is 
4.5$\times$10$^{29}$ erg s$^{-1}$, and that results in 
$\log(\frac{L_X^{unabs}}{L_{BOL}})=-5.6$. 
On the contrary, the large absorbing 
column, the large HRs, and the good fit by a power law of  
XMMU J000445.1+633550 (\#12),  XMMU J000457.6+634211 (\# 14) and
XMMU J000701.7+634151 (\# 54) suggest that these 3 sources may actually 
be background objects. The case of XMMU J000511.7+634018 (\# 19) and
XMMU J000537.1+634027 (\# 23) is different: they are also better fitted by 
a power law, but their absorbing column is small and their fluxes seem
rather low for close X-ray binaries.

\section{Conclusion}

We have obtained an \x\ observation of \hd\ and its surroundings. 
The peculiar star \hd\ was found to present a two-temperature spectrum,
and did not show any significant short-term variations during the 
exposure. These observations are also compatible with a stable X-ray 
emission since 1979. In parallel, we have continued our extensive optical 
monitoring of the star, and discovered that \hd\ continues to present 
dramatic line variations in the optical domain. The lack of significant 
changes in the X-ray emission compared to the optical data is a puzzle.
In such a context, no simple model is for the moment capable
of explaining the overall behaviour of one of the most peculiar 
O stars of the sky: any variation of the physical properties (magnetic 
field, mass-loss rate) of \hd\ is expected to have an impact on the 
whole wind, not only on the formation region of the optical emission 
but also on that of the X-rays. Long-term 
monitoring of \hd, preferentially in a multi-wavelength campaign, is 
thus necessary to eventually understand this star.

In addition, 57 new X-ray sources were also discovered
in the field of \hd, and we present here their characteristics (count 
rate, HRs, plus lightcurve analysis and spectral fits for the brightest 
ones). Only two correspond to rather bright stars, both of them being 
late-type stars with coronal X-ray emission. With large HRs and/or large 
absorbing columns and power law fits, several sources might be 
background objects. On the other hand, none of the 
B stars of the field was convincingly detected in X-rays.  

\begin{acknowledgements}
We acknowledge support from the PRODEX XMM-OM and Integral Projects 
and through contracts P4/05 and P5/36 `P\^ole d'Attraction Interuniversitaire' 
(Belgium). We are greatly indebted to the Fonds National de la Recherche Scientifique (Belgium) for multiple assistance including the financial support for the rent of the OHP telescope in 1999 and 2000 through contract 1.5.051.00 ``Cr\'edit aux Chercheurs'' FNRS. The travels to OHP for the observing runs were supported by the Minist\`ere de l'Enseignement Sup\'erieur et de la Recherche de la Communaut\'e Fran\c caise. 
The authors also thank the referee, Dr Pittard, for his useful comments.
This publication makes use of data products from the Two Micron All Sky Survey, USNO B1.0 and the Guide Star Catalog-II. 

\end{acknowledgements}


\begin{thebibliography}{99}
\bibitem[1989]{and} Anders, E., \& Grevesse, N. 1989, Geochimica et Cosmochimica Acta, 53,197
\bibitem[1985]{arn} Arnaud, M., \& Rothenflug, R. 1985, A\&AS, 60, 425
\bibitem[1997]{bab} Babel, J., \& Montmerle, T. 1997, ApJ, 485, L29
\bibitem[1997]{ber} Bergh\"ofer, T.W., Schmitt, J.H.M.M., Danner, R., \& Cassinelli, J.P. 1997, A\&A, 322, 167
\bibitem[1988]{bes} Bessell, M.S., \& Brett, J.M. 1988, PASP, 100, 1134
\bibitem[1989]{car} Cardelli, J.A., Clayton, G.C., \& Mathis, J.S. 1989, ApJ, 345, 245
\bibitem[1989]{chl} Chlebowski, T., Harnden, F.R.Jr., \& Sciortino, S. 1989, ApJ, 341, 427
\bibitem[1974]{con} Conti, P.S. 1974, ApJ, 187, 539
\bibitem[1973]{dav} Davidson, K.,  \& Ostriker, J.P. 1973, ApJ, 179, 585
\bibitem[2001]{RGS} den Herder, J.W., Brinkman, A.C., Kahn, S.M., et al.\ 2001, A\&A, 365, L7
\bibitem[1994]{dip} Diplas, A. \& Savage, B.D. 1994, ApJS, 93, 211
\bibitem[1997]{fel} Feldmeier, A., Kudritzki, R.-P., Palsa, R., Pauldrach, A.W.A., \& Puls, J. 1997, A\&A, 320, 899
\bibitem[1981]{fer} Ferrari-Toniolo, M., Persi, P.,\&  Grasdalen, G.L. 1981, PASP, 93, 633
\bibitem[2001]{gia} Giacconi, R., Rosati, P., Tozzi, P., et al. 2001, ApJ, 551, 624
\bibitem[1987]{gie} Gies, D.R.\ 1987, ApJS, 64, 545
\bibitem[1994]{har} Harris, D.E., Forman, W., Gioa, I.M., et al., EINSTEIN Observatory catalog of IPC X-ray sources (2E catalog),  SAO HEAD CD-ROM Series I ($Einstein$), Nos 18-36 (1994)
\bibitem[1977]{hat} Hatchett, S., \& McCray, R. 1977, ApJ, 211, 552
\bibitem[1989]{how89} Howarth, I.D., \& Prinja, R.K. 1989, ApJS, 69, 527
\bibitem[1997]{how97} Howarth, I.D., Siebert, K.W., Hussain, G.A.J., \&
 Prinja, R.K. 1997, MNRAS, 284, 265
\bibitem[1980]{hut} Hutchings, J.B., \& von Rudloff, I.R. 1980, ApJ, 238, 909
\bibitem[1997]{jef} Jeffries, R.D., Thurston, M.R., \& Pye, J.P. 1997, MNRAS, 287, 350
\bibitem[1992]{kaa92} Kaastra, J.S. 1992, An X-Ray Spectral Code for Optically Thin Plasmas (Internal SRON-Leiden Rep., Version 2.0) 
\bibitem[1993]{kaa93} Kaastra, J.S., \& Mewe, R. 1993, A\&AS, 97, 443
\bibitem[2002]{kaa02} Kaastra, J.S., Mewe, R., \& Raassen, A.J.J.\ 2003, in {\it New Visions of the X-ray Universe in the XMM-Newton and Chandra Era}, ed.\ F.\ Jansen, ESA SP-488, in press
\bibitem[2001]{Kahn} Kahn, S.M., Leutenegger, M.A., Cottam, J., et al.\ 2001, A\&A, 365, L312
\bibitem[1982]{kal} Kallman, T.R., \& McCray, R. 1982, ApJS, 50, 263
\bibitem[1995]{ken} Kenyon, S.J., \& Hartmann, L. 1995, ApJS, 101, 117
\bibitem[1984]{kro} Krolik, J.H., \& Kallman, T.R. 1984, ApJ, 286, 366
\bibitem[1996]{kud} Kudritzki, R.P., Palsa, R., Feldmeier, A., Puls, J., \& Pauldrach, A.W.A. 1996, Proc. 'R\"ontgenstrahlung from the Universe', eds. Zimmermann, H.U., Tr\"umper, J., and Yorke, H.;, MPE Report 263, 9
\bibitem[1993]{lam93} Lamers, H.J.G.L.M., \& Leitherer, C. 1993, ApJ, 412, 771
\bibitem[1994]{lam94} Lamers, H.J.G.L.M., \& Morris, P. 1994, private communication
\bibitem[1973]{mih} Mihalas, D. 1973, PASP, 85, 593 
\bibitem[2002]{mof} Moffat, A.F.J., Corcoran, M.F., Stevens, I.R., et al. 2002, ApJ, 573, 191
\bibitem[2000]{mus} Mushotzky, R.F., Cowie, L.L., Barger, A.J., \& Arnaud, K.A. 2000, Nature, 404, 459
\bibitem[2001]{naz} Naz\'e, Y., Vreux, J.-M., \& Rauw, G.  2001, A\&A, 372, 195 (Paper I)
\bibitem[1999]{owo} Owocki, S.P., \& Cohen, D.H. 1999, ApJ, 520, 833
\bibitem[1984]{pep} Peppel, U. 1984, ApJS, 56, 257
\bibitem[2002a]{9sgr} Rauw, G., Blomme, R., Waldron, W.L., et al.\ 2002a, A\&A, 394, 993
\bibitem[2002b]{rau} Rauw, G., Naz\'e, Y., Gosset, E., et al. 2002b, A\&A, 395, 499
\bibitem[2000]{rxp} \ro\ Consortium, The Second \ro\ Source Catalog of Pointed Observations (2RXP), \ro\ News 72, 25-May-2000 (2000)
\bibitem[2000]{rxh} \ro\ Scientific Team, The \ro\ Source Catalog of Pointed Observations with the High Resolution Imager (1RXH) (3rd Release), \ro\ NEWS No. 71, The \ro\ Consortium (2000)
\bibitem[2003]{san} Sana, H., Stevens, I.R., Gosset, E., Rauw, G., \& Vreux, J.-M. 2003, MNRAS, submitted
\bibitem[1998]{schle} Schlegel, D.J., Finkbeiner, D.P., \& Davis, M. 1998, ApJ, 500, 525
\bibitem[1982]{sch} Schmidt-Kaler, T. 1982, Land\"olt-Bornstein Catalogue VI/2b
\bibitem[1982]{shu} Shull, J.M., \& Van Steenberg, M. 1982, ApJS, 48, 95
\bibitem[1982]{shuerr} Shull, J.M., \& Van Steenberg, M. 1982, ApJS, 49, 351 (errata)
\bibitem[1996]{sin} Singh, K.P., White, N.E., \& Drake, S.A. 1996, ApJ, 456, 766
\bibitem[1986]{ste} Stella, L., White, N.E., \& Rosner, R. 1986, ApJ, 308, 669
\bibitem[2001]{str} Str\"uder, L., Briel, U., Dennerl, K. et al. 2001, A\&A, 365, L18
\bibitem[1948]{swi} Swings, P. 1948, An. Astrop., 11, 228 
\bibitem[1969]{tar} Tarter, C.B., Tucker, W.H., \& Salpeter, E.E. 1969, ApJ, 156, 943
\bibitem[2001]{tur} Turner, M.J.L., Abbey, A., Arnaud, M., et al. 2001, A\&A, 365, L27
\bibitem[2002]{ud} ud-Doula, A. 2002, PhD thesis, University of Delaware
\bibitem[1995]{ver95} Verner, D.A., \& Yakovlev, D.G. 1995, A\&AS, 109, 125
\bibitem[1996]{ver96} Verner, D.A., Ferland, G.J., Korista, K.T., \& Yakovlev, D.G. 1996, ApJ, 465, 487
\bibitem[1996]{verfer} Verner, D.A., \& Ferland, G.J. 1996, ApJS, 103, 467
\bibitem[1999]{vog99} Voges, W., Aschenbach, B., Boller, Th., et al. 1999, A\&A, 349, 389 (1RXS catalog)
\bibitem[2000]{vog00} Voges, W., Aschenbach, B., Boller, Th., et al. 2000, \ro\ All-Sky Survey Faint source Catalogue (RASS-FSC), Max-Planck-Institut f\"ur extraterrestrische Physik, Garching 
\bibitem[1997]{vor} Voronov, G.S. 1997, Atomic Data and Nuclear data Tables, 65, 1
\bibitem[2003]{wal03} Walborn, N.R., Howarth, I.D., Herrero, A., \& Lennon, D.J. 2003, ApJ, 588, 1025
\bibitem[1984]{wal84} Waldron, W.L. 1984, ApJ, 282, 256
\bibitem[2000]{wga} White, N.E., Giommi, P., \& Angelini, L. 2000, The WGACAT version of the \ro\ PSPC Catalogue, Rev. 1, Laboratory for High Energy Astrophysics (LHEA/NASA), Greenbelt 
\end{thebibliography}
\end{document}